\documentclass[twocolumn,secnumarabic,showpacs,preprintnumbers,nobibnotes,showfootnotes,amssymb,aps,prl]{revtex4}

\usepackage{graphicx}
\usepackage{dcolumn}
\usepackage{bm}
\usepackage{epsfig}
\newcommand{\newc}{\newcommand}

\def\beq{\begin{equation}}
\def\eeq{\end{equation}}
\def\bea{\begin{eqnarray}}
\def\eea{\end{eqnarray}}

\newc{\ie}{{\it i.e.~}}          \newc{\etal}{{\it et al.~}}
\newc{\eg}{{\it e.g.~}}          \newc{\etc}{{\it etc.~}}
\newc{\cf}{{\it c.f.~}}
\newc{\gsim}{\lower.7ex\hbox{$\;\stackrel{\textstyle>}{\sim}\;$}}
\newc{\lsim}{\lower.7ex\hbox{$\;\stackrel{\textstyle<}{\sim}\;$}}
\newc{\gev}{\,{\rm GeV}}
\newc{\mev}{\,{\rm MeV}}
\newc{\ev}{\,{\rm eV}}
\newc{\kev}{\,{\rm keV}}
\newc{\tev}{\,{\rm TeV}}

\newc{\gst}{\hat{g}_3}
\newc{\gw}{\hat{g}_2}
\newc{\gy}{\hat{g}_Y}
\newc{\g}{\hat{g}}
\newc{\muh}{\hat{\mu}}
\newc{\lambdah}{\hat{\lambda}}

\begin{document}

\preprint{MCTP-09-37}
\preprint{FERMILAB-PUB-09-384-A}

\title{Kaluza-Klein Dark Matter And Neutrinos From Annihilation In The Sun}

\author{Thomas Flacke$^{1,2}$}

\author{Arjun Menon$^1$}

\author{Dan Hooper$^{3,4}$}

\author{Katherine Freese$^{1}$}

\affiliation{$^1$ Michigan Center for Theoretical Physics (MCTP),  
Department of Physics, \\ University of Michigan, Ann Arbor, MI 48109 USA}
\affiliation{$^2$ Institut f\"ur Theoretische Physik und Astrophysik, 
Universit\"at  W\"urzburg, D-97074 W\"urzburg, Germany}
\affiliation{$^3$ Center for Particle Astrophysics, Fermi National Accelerator 
Laboratory, Batavia, IL 60510 USA}
\affiliation{$^4$ Astronomy and Astrophysics Department, University of Chicago, 
Chicago, IL 60637 USA}


     

\begin{abstract}
In models with one universal extra dimension (UED), the first Kaluza-Klein excitations of the hypercharge gauge boson, $B^{(1)}$, and the neutral component of isospin gauge boson, $W^{3(1)}$, are each viable dark matter candidates. In either case, such particles are predicted to accumulate in the core of the Sun, where they annihilate to generate a potentially observable flux of high energy neutrinos. In this article, we calculate the flux of neutrinos produced in this model and determine the constraints that can be placed on the UED parameter space from current IceCube data. For the case of $B^{(1)}$ dark matter, we find that the present limits from IceCube are stronger than those from direct dark matter detection experiments such as CDMS and XENON10. For $W^{3(1)}$ dark matter, the present IceCube data provides a constraint slightly weaker than direct detection experiments. In addition, we also present the projected regions of UED parameter space that can be probed by IceCube/DeepCore in the near future and compare them to the prospects for future direct detection experiments.
\end{abstract}

\date{\today}

\maketitle



\section{Introduction}
In models with one universal extra dimension (UED)~\cite{Appelquist:2000nn}, all of the Standard Model particles are promoted to 5 dimensional fields, propagating in a flat extra dimension (for earlier ideas closely related to UED models see Ref.~\cite{preUED}). In order to allow for the existence of chiral zero-mode fermions, the extra dimension is modded out by the orbifold, $S^1/\mathbb{Z}_2$. Due to the $\mathbb{Z}_2$ symmetry, the interactions between Kaluza-Klein (KK) modes respect what is known as KK-parity, which implies that Kaluza-Klein particles can only
be produced or destroyed in pairs. Among other consequences, this leads to a lower bound on the KK mass scale of only $M_{KK}\gsim 300 \gev$ \cite{UEDbounds}. Furthermore, KK-parity guarantees the stability of the lightest Kaluza-Klein particle (LKP), thus providing us with a potentially viable dark matter candidate. For a review of the UED model and its phenomenology, see Ref.~\cite{Hooper:2007qk}.

Despite its minimal field content, the UED model contains a 
large number of undetermined parameters. In addition to the Higgs 
mass, $m_h$, and the compactification radius, $R$, the geometry also allows for 
operators of the 5D fields at the orbifold fixed points whose magnitudes are
independent of the bulk parameters (conservation of KK-parity forces 
the boundary localized operators to be included symmetrically on both fixed 
points, however). In the absence of such boundary localized terms, one expects the first KK excitation of the hypercharge gauge boson, $B^{(1)}$, to be the LKP~\cite{servanttait}. The inclusion of such terms, however, can affect the KK mass spectrum~\cite{BLKTrefs}, and allow for other states to potentially serve as the LKP. As shown in Ref.~\cite{Flacke:2008ne}, the first KK excitation of the neutral $SU(2)$ gauge boson, $W^{3(1)}$, can be the LKP if small but non-vanishing boundary kinetic terms in the electroweak sector are considered. Other identities for the LKP are more difficult to realize. The first KK excitation of 
the Higgs boson, $h^{(1)}$, could be the LKP, but would require large boundary 
localized terms and is strongly constrained for low $h^{(1)}$ masses~\cite{Flacke:2008ne}.  No experimentally viable realizations are known in which the LKP takes the form of a KK-pseudoscalar Higgs or KK-neutrino~\cite{Flacke:2008ne,ST2}.


The hypothesis that dark matter takes the form of stable, weakly interacting massive particles (WIMPs), is gradually being tested by direct and indirect detection experiments. At present, 
the strongest direct detection constraints on the spin-independent scattering 
cross section of WIMPs off nuclei have been placed by the CDMS~\cite{Ahmed:2008eu} and 
XENON~\cite{Angle:2007uj} collaborations (the DAMA collaboration has also reported an annual modulation of their rate which they interpret as a positive detection of dark matter~\cite{dama}). In addition, WIMPs can produce potentially observable particles in their annihilations. WIMPs which scatter with nuclei in the Sun can be captured and accumulate in the center of Sun, where they ultimately annihilate to produce high energy neutrinos
that can be observed with large volume neutrino telescopes, such as IceCube~\cite{Abbasi:2009uz}. Assuming that the time scale for the annihilation and capture rates to equilibrate in the Sun is smaller than the age of the Solar System, the present annihilation rate of WIMPs will be determined only by the capture rate of WIMPs which, in turn, can be determined by the WIMP-nucleon scattering cross section~\cite{Jungman:1995df}. Therefore both direct detection experiments and neutrino telescopes are sensitive to similar elastic scattering processes.

In this article, we study the impact of these indirect and direct detection
constraints on the UED parameter space for LKPs in the form of a $B^{(1)}$ or $W^{3(1)}$. The phenomenology of a $B^{(1)}$ LKP has been studied 
extensively~\cite{servanttait,CFM,UEDDMrefs}, including the prospects for direct~\cite{CFM,ST2,AKM} and indirect detection~\cite{CFM,kkindirect,HK}. The analysis for detecting $B^{(1)}$ dark matter via high energy neutrinos from WIMP annihilation in the Sun was first performed in 
Ref.~\cite{HK}, which we closely follow in this article. The thermal relic abundance of a $W^{3(1)}$ LKP, and the related constraints from direct detection experiments, have been calculated in Ref.~\cite{AKM} which, to the best of our knowledge, is the only study to consider this dark matter candidate (for discussion of dark matter phenomenology in models with two universal extra dimensions, see Ref.~\cite{6D}).

In this article, we apply the analysis of Ref.~\cite{HK} to the cases of $B^{(1)}$ and $W^{3(1)}$ dark matter. In each case, the LKPs scatter off nucleons through the $t$-channel exchange of KK-fermions. The experimental limits from direct detection experiments and IceCube
lead to strong constraints on the $m_{DM} \-- r_q$ plane, where $m_{DM}$ is the
LKP mass and $r_q$ is the fractional splitting between the LKP mass and that of the KK-quarks. The constraints that derive from CDMS and XENON lead to limits on 
the WIMP-nucleon spin-independent cross sections, while those from IceCube 
primarily lead to bounds on the WIMP-nucleon spin-dependent cross sections.





\section{Elastic Scattering of $B^{(1)}$ and $W^{3(1)}$ Dark Matter}

Following Ref.~\cite{Jungman:1995df}, the spin-independent dark matter-nuclei elastic scattering 
cross section is given by 
\bea
\sigma_{SI} = \frac{m_T^2}{4\pi (m_{DM}+m_T)^2} \left[Z f_p+(A-Z)f_n
\right]^2,
\label{dmSIsigma}
\eea 
where $m_{DM}$ is the WIMP mass, $m_T$ is the mass of the target nucleus, $Z$ 
is the number of protons in the nucleus, and $A$ is the number of nucleons in the 
nucleus. The WIMP-nucleon couplings, $f_{p,n}$, can be written in terms of the couplings to individual quarks:
\bea
f_{p,n} = \sum_{q=u,d,s} \frac{\beta_q}{m_q} m_{p,n} f_{T_q}^{p,n}, \label{fpn}
\eea
where $m_{p,n}$ are the respective neutron and proton mass, and the nucleon matrix 
elements are $f_{T_u}^p=0.020\pm0.004$, $f_{T_d}^p=0.026 \pm 0.005$, $f_{T_u}^n=
0.014\pm0.003$, $f_{T_d}^n=0.036 \pm 0.008$ and $f_{T_s}^{p,n}=0.118 \pm 0.062
$~\cite{dmdetect}. Following Ref.~\cite{AKM}, we have conservatively neglected the couplings to gluons through heavy quark loops. The quark-level couplings for a $B^{(1)}$ LKP are given (in the limit of small mixing between left- and right-handed KK-fermions and small splittings between the KK-fermions mass and the LKP mass) by~\cite{ST2}
\bea
\beta_{q} \approx  \frac{e^2m_q}{2m_{B^{(1)}}^2\cos^2 \theta_W} \left(
\frac{Y_{q_L}^2}{r_{q_L}^2} + \frac{Y_{q_R}^2}{r_{q_R}^2}\right)
\eea
or, for a $W^{3(1)}$ LKP, by
\bea
\beta_{q} \approx  \frac{e^2m_q}{8 m_{W^{3(1)}}^2\sin^2 \theta_W} \left(\frac{1}{r_{q_L}^2}\right)
\eea
where 
\beq
r_{q_{L,R}}\equiv \frac{m_{q^{(1)}_{L,R}}-m_{DM}}{m_{DM}},
\label{rqdef}
\eeq
$e$ is the electric charge, $\theta_W$ is the Weinberg angle, $Y_{q_{L,R}}$ 
are the appropriate hypercharges, and $m_{q_{L,R}^{(1)}}$ are the left- and 
right-handed KK-quark masses. From this, it is evident that the scattering
cross section is greatly enhanced in the case of small mass splittings.

Although direct detection experiments are currently most sensitive to spin-independent scattering, spin-dependent scattering can in many cases dominate the capture rate of WIMPs in the Sun. Within the context of UED, in particular, spin-dependent scattering typically dominate this process. This is, in particular, true for the dark matter candidates considered in this article.
%
%
%
%
%
%
%
The spin dependent 
scattering off protons ({\it ie.} hydrogen nuclei) arises from the $t$-channel exchange of KK-quarks and is given by~\cite{CFM,AKM}
\bea
\sigma^{\rm H}_{{\rm SD},B^{(1)}}&=&\frac{g_1^4 m^2_p}{648 \pi m^4_{B^{(1)}}r^2_{q_R}}(4\Delta^p_u+
\Delta^p_d+\Delta^p_s)^2 \nonumber \\
&\approx& 1.7 \times 10^{-6} {\rm pb} \, \bigg(\frac{1 \,{\rm TeV}}{m_{B^{(1)}}}\bigg)^4 \bigg(\frac{0.1}{r_{q_R}}\bigg)^2, \\
\sigma^{\rm H}_{{\rm SD},W^{3(1)}}&=&\frac{g_2^4 m^2_p}{128 \pi m^4_{W^{3(1)}}r^2_{q_L}}(\Delta^p_u+
\Delta^p_d+\Delta^p_s)^2 \label{ESXW} \nonumber\\
&\approx& 0.36 \times 10^{-6} {\rm pb} \, \bigg(\frac{1 \,{\rm TeV}}{m_{W^{3(1)}}}\bigg)^4 \bigg(\frac{0.1}{r_{q_L}}\bigg)^2,
\label{ESXB}
\eea
where $\Delta^p_u=0.78\pm 0.02$, $\Delta^p_d=-0.48\pm 0.02$, and $\Delta^p_s=-0.15\pm 0.07$
quantify the spin-content of the quarks within the proton~\cite{delta}. As 
only left-handed quarks couple to $W^{3(1)}$, $\sigma^{\rm H}_{SD,W^{3(1)}}$ depends 
only on $r_{q_L}$. For $B^{(1)}$, both $q_L^{(1)}$ and 
$q_R^{(1)}$ contribute, but the cross section is dominated by $q_R^{(1)}$ due to the larger $U(1)_Y$ charges of right-handed fermions.
Therefore in Eq.~(\ref{ESXB}) we neglect the $q_L^{(1)}$ contributions.
In addition we would also like to emphasize that for $\sigma^{\rm H}_{SD,W^{3(1)}}$ the 
spin-content is given by $(\Delta^p_u+\Delta^p_d+\Delta^p_s)^2=.023$, where as
for $\sigma^{\rm H}_{SD,B^{(1)}}$ it is of $\mathcal{O}(1)$. Thus despite the fact that the $g_2$ appearing in $\sigma^{\rm H}_{SD,W^{3(1)}}$ is about twice as large as the $g_1$ in 
$\sigma^{\rm H}_{SD,B^{(1)}}$, $\sigma^{\rm H}_{SD,W^{3(1)}}$ is suppressed relative to $\sigma^{\rm H}_{SD,B^{(1)}}$.



\section{Capture and Annihilation In The Sun}

Using the scattering cross sections described in the previous section, we can estimate the rate at which LKPs will be captured in the Sun.  Assuming a local dark matter density of 0.3 GeV/cm$^3$, the rate of WIMPs captured in the Sun is given by~\cite{Jungman:1994jr}
\beq
C_{\odot}\approx(3.35 \times 10^{18} \, {\rm s}^{-1})
\left(\frac{270 \, {\rm km/s}}{\bar{v}} \right)^3
\left(\frac{\sigma_{\rm Eff}}{10^{-6}\,{\rm pb}}\right)
\left(\frac{1 \tev}{m_{DM}}\right)^2
\eeq
where $\bar{v}$ is the local RMS velocity of the dark matter halo distribution and $\sigma_{\rm Eff}$ is the effective elastic scattering cross section, given by
\begin{eqnarray}
\sigma_{\rm Eff} &\approx& \sigma^{\rm H}_{\rm SD} + \sum_i F_i A_i K_i(m_{\rm DM}) \sigma^{\rm i}_{\rm SI} \nonumber \\
&\approx& \sigma^{\rm H}_{\rm SD} + \sigma^{\rm H}_{\rm SI} + 0.175 \sigma^{\rm He}_{\rm SI} + ...
\end{eqnarray}
In this expression, $F_i$, $A_i$ and $K_i(m_{DM})$ denote the relative degree of form factor suppression, solar abundances, and kinematic suppression for the various species of nuclei. These factors are evaluated for the case of hydrogen and helium nuclei in the lower expression. Spin-independent scattering with oxygen nuclei can also be important and may even dominate over spin-independent scattering with other nuclei for sufficiently heavy WIMPs. Any spin-independent scattering cross section that is not already excluded by direct detection experiments, however, will generate an unobservably small flux of high energy neutrinos~\cite{halzen}. Therefore, we focus throughout the remainder of this article on WIMP capture in the Sun through spin-dependent scattering, which is far less stringently constrained.

For the range of masses, elastic scattering cross sections, and annihilation cross sections we consider in this paper, we have explicitly checked that the processes of dark matter capture and annihilation in the Sun reach equilibrium~\cite{Jungman:1994jr}. In all cases shown, the rate of WIMPs annihilating in the Sun is very close to the rate of WIMPs captured. This is in contrast to WIMP capture and annihilation in the Earth, which does not reach equlibrium in the amount of time available~\cite{earth}.

To calculate the flux of neutrinos from the Sun, we need to know the 
fraction of LKP annihilations which proceed to various Standard Model states. In Table~\ref{tab:BR}, we show the annihilation ratios for $B^{(1)}$ and 
$W^{3(1)}$ LKPs annihilations. For the $B^{(1)}$, annihilations occur largely through the 
exchange of KK-quarks and leptons, and the fraction to each final state is determined by the $U(1)_Y$ charges of the exchanged and final state particles. In contrast, the $W^{3(1)}$ dominantly 
annihilates to $W^+W^-$ via the $t$-channel exchange of a KK-$W^{\pm}$.

\begin{table}
\begin{center}
\begin{tabular}{|c|c|c|c|c|c|c|c|}
\hline
 &$W^+W^-$&$\bar{\nu}\nu$&$\tau^+\tau^-$&$\bar{t}t$&$\bar{b}b$&$\bar{c}c$&$\bar{q}q$\\
\hline
$B^{(1)}$ & 0  &0.04&0.20&0.11&0.01&0.11&0.12\\
\hline
$W^{3(1)}$& 0.65&0.04&0.01&0.04&0.04&0.04&0.13\\
\hline
\end{tabular}
\end{center}
\caption{The approximate fraction of $B^{(1)}$ and $W^{3(1)}$ annihilations which proceed to various Standard Model final states. The $\bar{\nu}\nu$ channel is summed over all neutrino species. The $\bar{q}q$ channel is summed over $u,d,$ and $s$ quarks.}
\label{tab:BR}
\end{table}

\begin{figure}[t]
\includegraphics[width=0.48\textwidth]{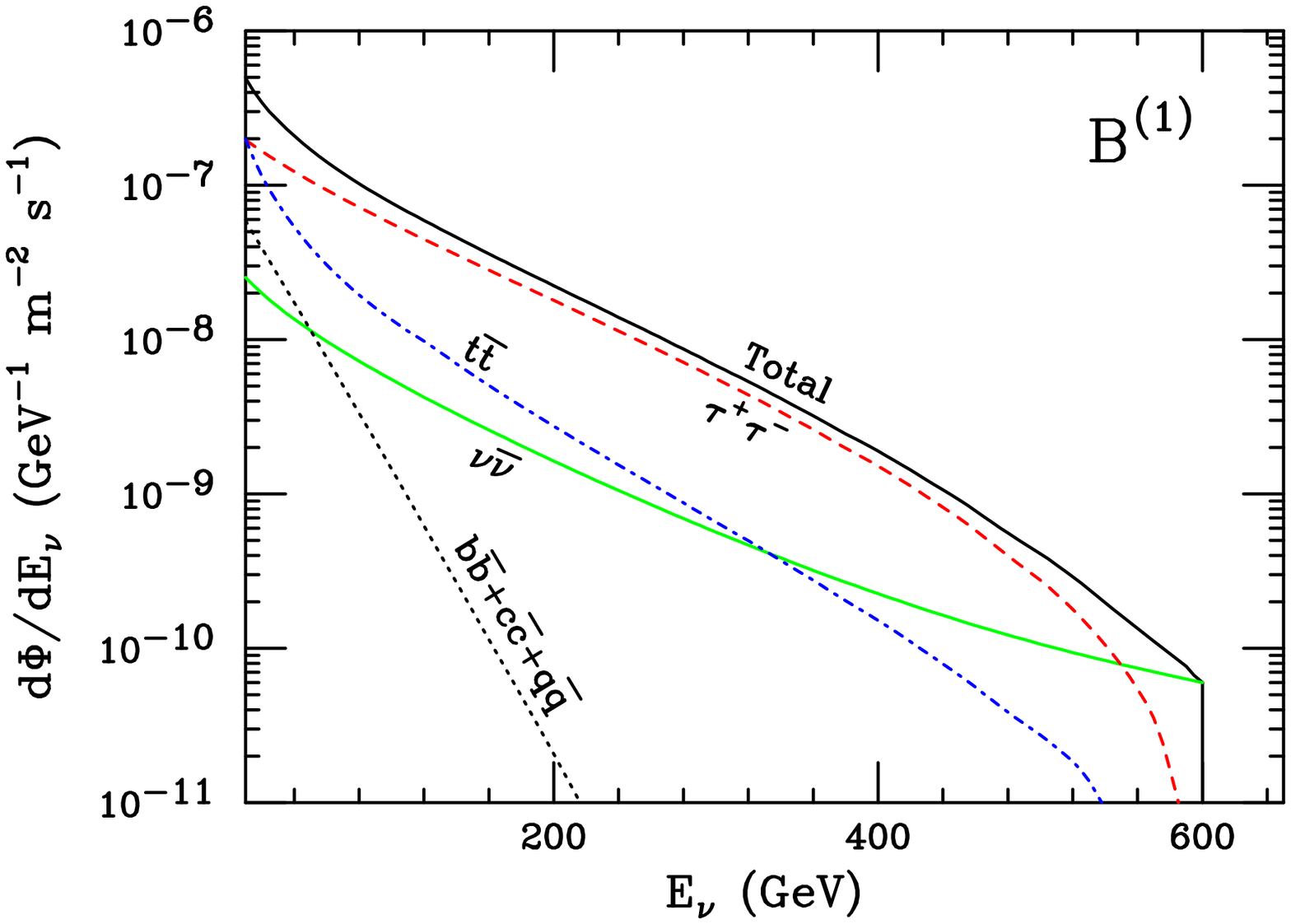}\\
\includegraphics[width=0.48\textwidth]{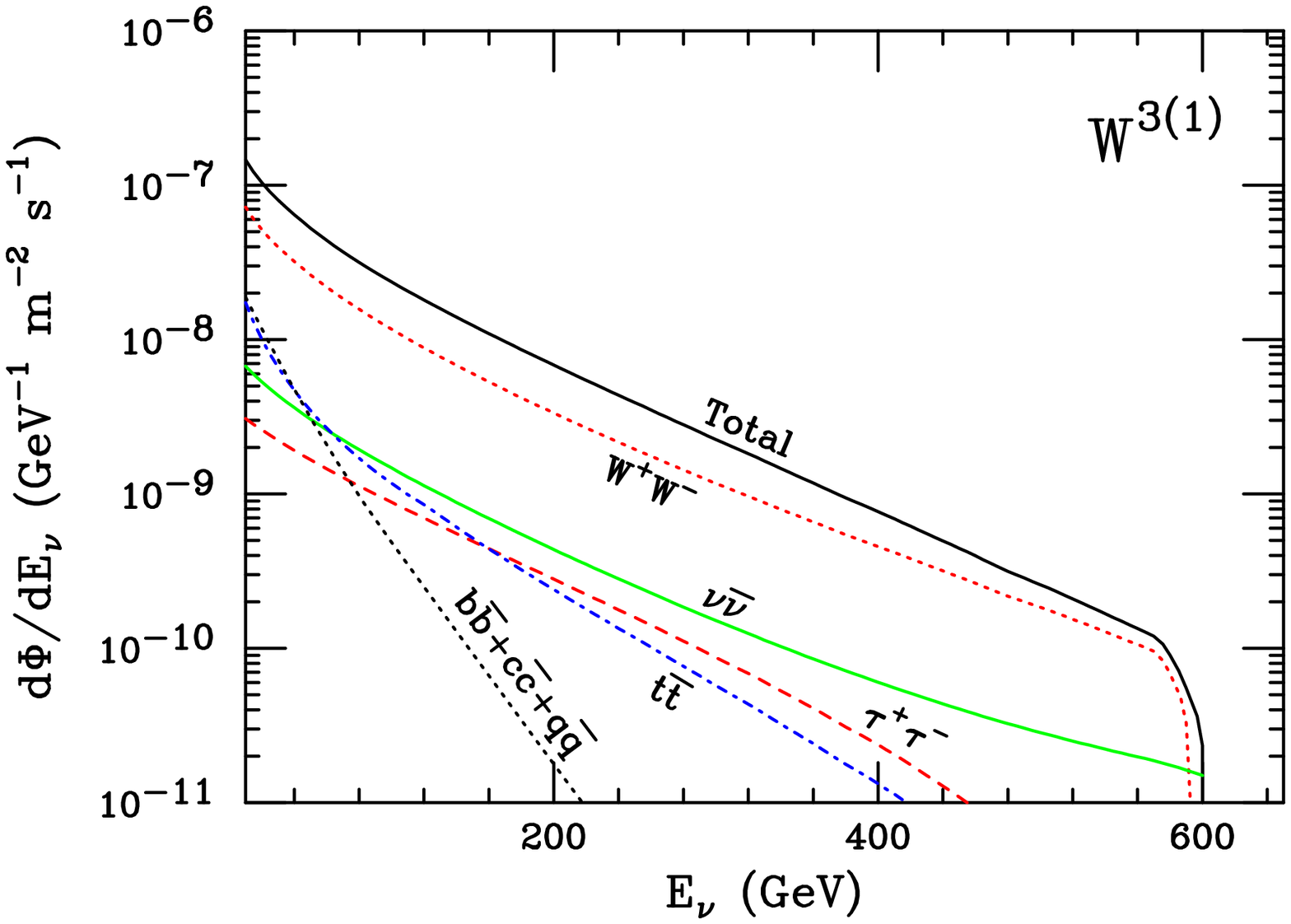}
\caption{The spectrum of high energy neutrinos from $B^{(1)}$ (top) or $W^{3(1)}$ (bottom) dark matter annihilations in the Sun. In each frame, we also show the contributions to the neutrino spectrum from various annihilation channels. In the case of $B^{(1)}$ dark matter, the neutrino flux is dominated by annihilations to taus and neutrinos, whereas for $W^{3(1)}$ dark matter, annihilations to $W^+ W^-$ produce the most neutrinos. In each case we have used $r_{q_{L,R}}=0.14$ and $m_{DM}=600 \gev$.}
\label{fig:channels}
\end{figure}

These ratios determine the injection spectrum of high energy 
neutrinos in the Sun. Using this injection spectrum we calculate the $\nu_\mu$ 
energy spectrum at the Earth using the tabulated results of Ref.~\cite{Cirelli:2005gh}. In 
Fig.~\ref{fig:channels} we present the flux of high energy neutrinos, 
$d\Phi/dE_\nu$,
due to annihilation of $B^{(1)}$ or $W^{3(1)}$ LKPs in the Sun for 
$m_{DM}=600$~GeV. Neutrinos coming from $\tau$'s dominate the neutrino spectrum for the $B^{(1)}$ case, while for $W^{3(1)}$ dark matter the dominant contribution is from $W^+W^-$. The neutrino spectrum from both 
of these dark matter candidates is generally hard, enabling us to use the more stringent IceCube limits, as opposed to the less stringent limits for annihilations to $b\bar{b}$~\cite{Abbasi:2009uz}. In the case of $B^{(1)}$ annihilations, the spectrum is actually harder than those considered in the IceCube analysis, making the results presented here conservative. As
electroweak precision constraints require the compactification scale to be greater
than $\approx 300$~GeV, we do not consider the constraints from Super Kamiokande which apply mostly to the case of low mass WIMPs.


\section{Current Limits From And Future Prospects For IceCube}

\begin{figure}[t]
\includegraphics[width=0.48\textwidth]{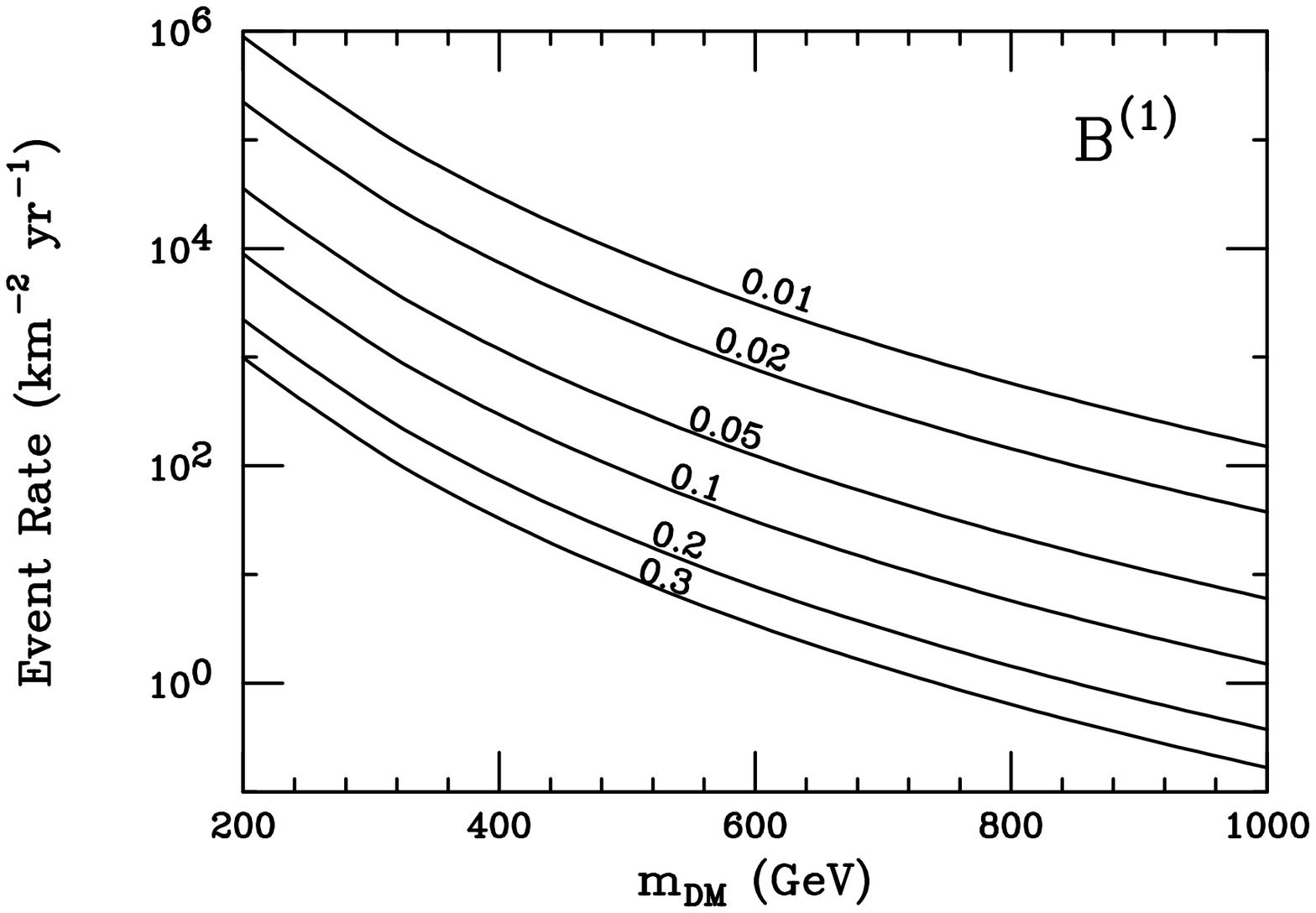}
\includegraphics[width=0.48\textwidth]{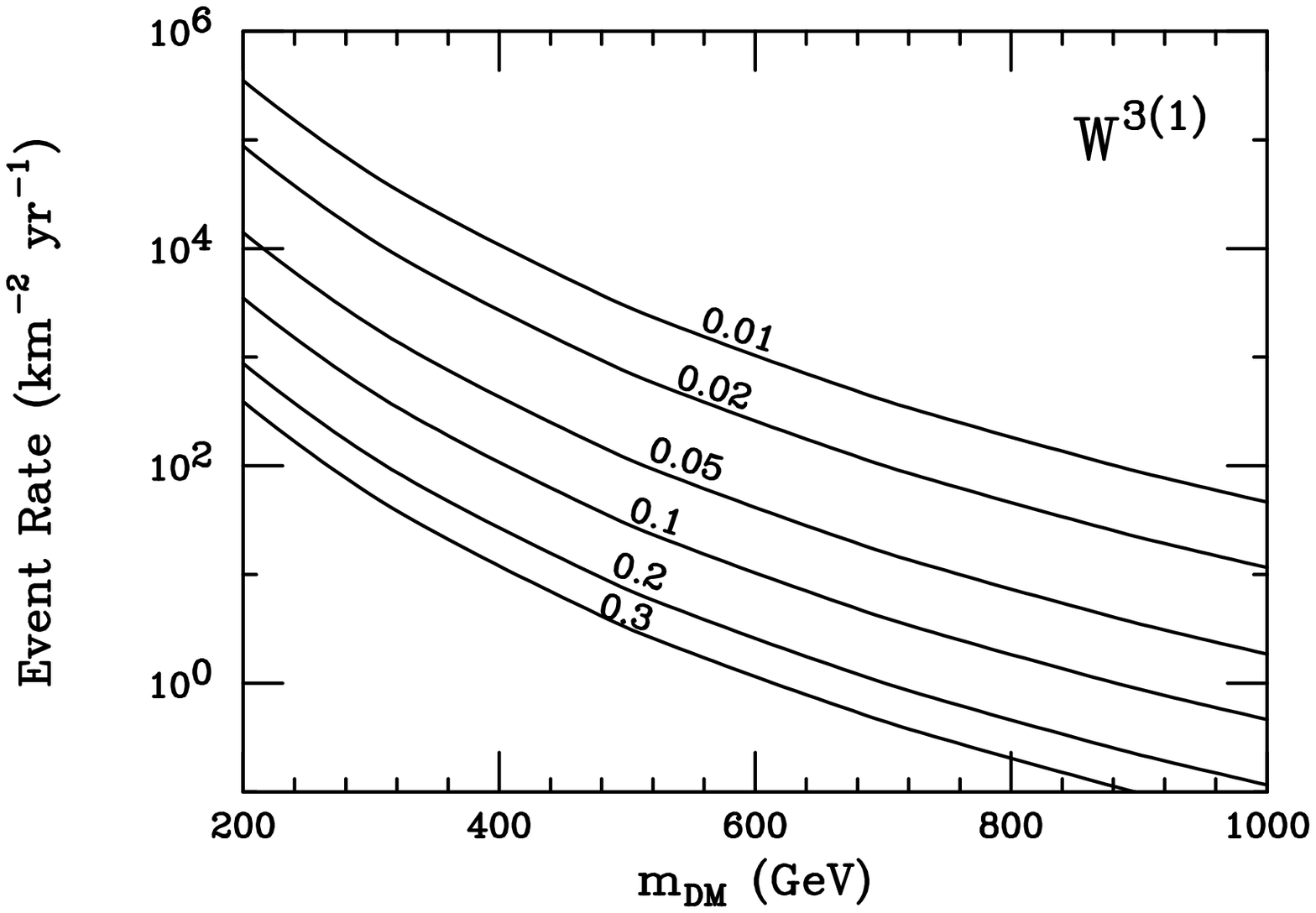}
\caption{The rate of neutrino-induced muons per effective area, in a kilometer-scale neutrino telescope such as IceCube, from $B^{(1)}$ (top) or $W^{3(1)}$ (bottom) dark matter annihilations in the Sun. The six contours shown in each frame correspond to mass splittings between the LKP and the KK-quarks of $r_{q_{L,R}}=$0.01, 0.02, 0.05, 0.1, 0.2 and 0.3, from top-to-bottom.}
\label{fig:detrate}
\end{figure}

To calculate the rate of neutrino induced muon tracks from WIMP annihilations in a large volume neutrino telescope such as IceCube, we use the following expression:
\bea
R= \int dE_{\nu} dy A_{\rm eff} N_A \bigg[\frac{d\sigma_{\nu N}}{dy} \frac{d\Phi}{dE_{\nu}} + (\nu \to \bar{\nu}) \bigg] [R_\mu(E_\mu)+D], \nonumber 
\label{nevnts}
\eea
\vspace{-0.7cm}
\bea
\eea
where $A_{\rm eff}$ is the effective area of the detector, $d\sigma_{\nu N}/dy$ is the differential neutrino-nucleon charged current cross section, $1-y$ is the fraction of the neutrinos's energy that is transferred to the muon in the interaction, $D$ is the depth of the detector ($\sim 1$ km for horizontal muons in IceCube), and $R(E_{\mu})$ is the muon range which is approximately given by~\cite{range}
\begin{equation}
R_{\mu} (E_\mu) = \frac{1}{\rho \beta} \log \left(\frac{ \alpha + \beta E_{\mu}}{
\alpha + \beta E_{thresh}} \right).
\end{equation}
Here, $\rho$ is the density of ice, $\alpha \approx 2.0$ MeV cm$^{2}$ g$^{-1}
$, $\beta \approx 2.0$ MeV cm$^{2}$ g$^{-1}$ and we take the threshold energy
to be $E_{thresh}=50$~GeV, although lower values will likely be attained with the future inclusion of the DeepCore array within the larger IceCube volume~\cite{Resconi:2008fe}. The charged current neutrino and antineutrino-nucleon cross sections are given by~\cite{sigma}
\begin{eqnarray}
\frac{d\sigma_{\nu N}}{d E_{\mu}} = \frac{2}{\pi} m_{N} G_{F}^2 \left(a_{N \nu} + b_{N \nu} 
\frac{E_{\mu}^{2}}{E_{\nu}^2} \right),
\end{eqnarray}
where $a_{p \nu}=0.15$, $b_{p \nu}=0.04$, $a_{n \nu}=0.25$, $b_{n \nu}=0.06$, and 
the corresponding expressions for anti-neutrinos can be found by 
$a_{p \bar{\nu}}=b_{n\nu}$, $b_{p \bar{\nu}}=a_{n \nu}$, $a_{n \bar{\nu}}=b_{p \nu}$, 
$b_{n \bar{\nu}}=a_{p \nu}$. In Fig.~\ref{fig:detrate} we show the event
rate per square kilometer of effective area as function of LKP mass for both the $B^{(1)}$ and $W^{3(1)}$ LKPs, and for different fractional mass splittings, $r_q$. As expected smaller splittings and 
lower LKP masses lead to larger event rates at IceCube.

\begin{figure}[t]
\includegraphics[width=0.48\textwidth]{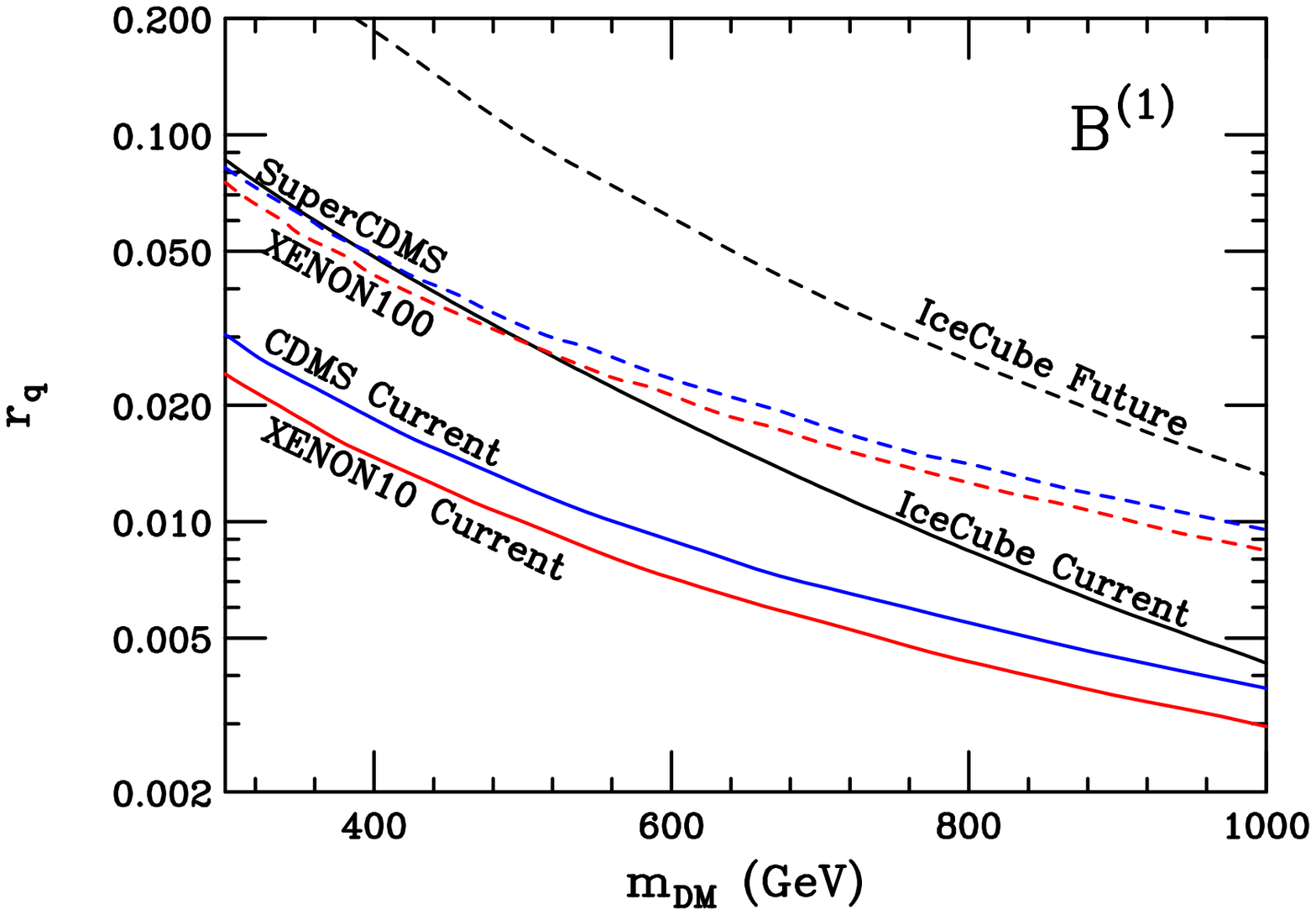}
\hspace{0.1cm}
\includegraphics[width=0.48\textwidth]{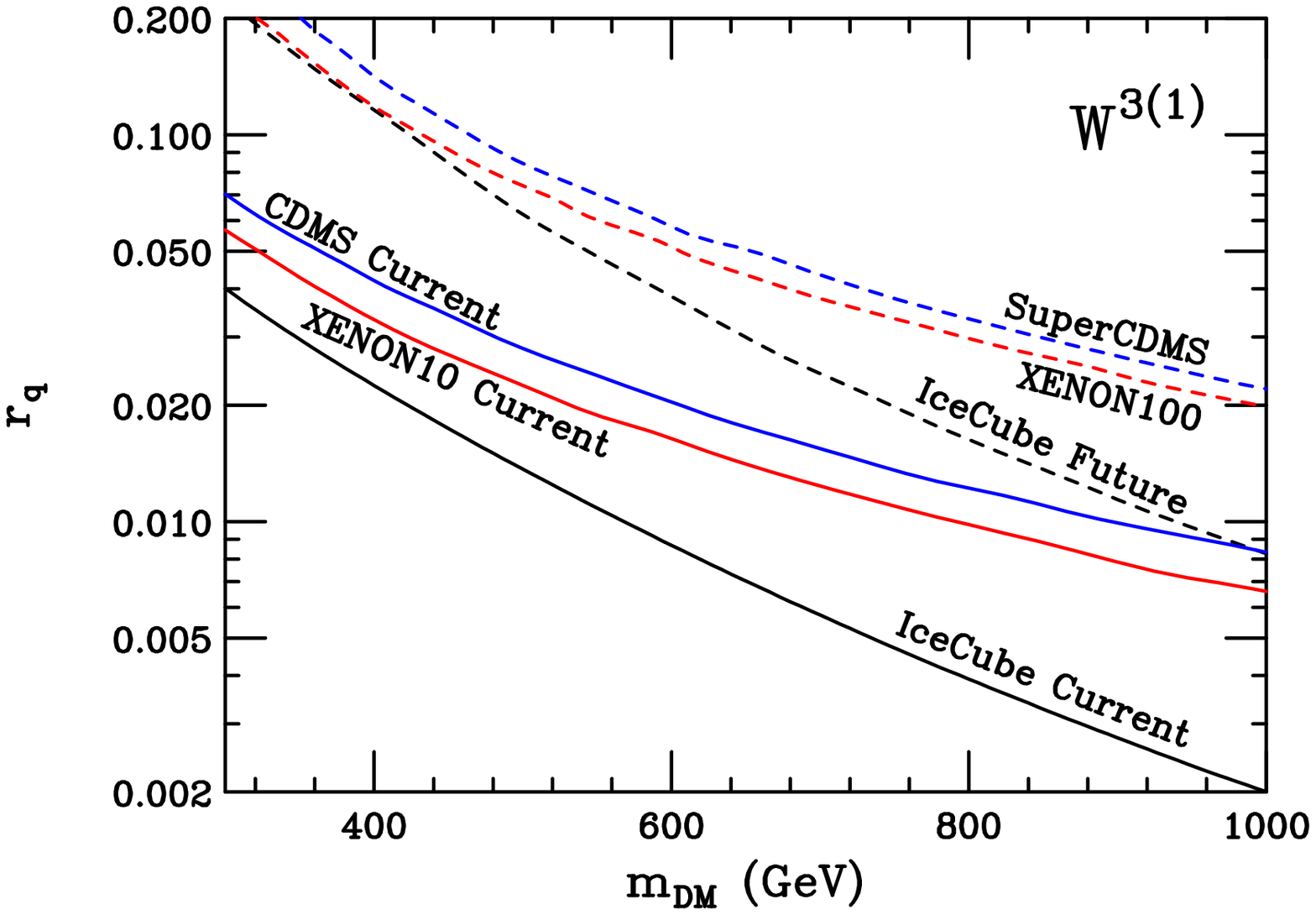}
\caption{The current limits (solid lines) and projected constraints (dashed lines) on the UED parameter space from IceCube and direct detection experiments. The top and bottom frames show the results for the cases in which the LKP is the $B^{(1)}$ (top) or the $W^{3(1)}$ (bottom), respectively. The regions below the solid lines are presently excluded by the corresponding
experimental limit, while the parameter space below the dashed lines is predicted to be 
probed by the corresponding experiment in the future.}
\label{fig:bounds}
\end{figure}

We next turn our attention to placing constraints on the UED parameter space from the existing IceCube data (from 2007, when IceCube consisted of 22 strings of detectors). IceCube's current limits on WIMP annihilations can be found in Ref.~\cite{Abbasi:2009uz}. Rather than being given in terms of event rates, the limits are given in terms of the dark matter annihilation rate, or the WIMP's elastic scattering cross section with nuclei (for two choices of the annihilation channel). In Fig.~\ref{fig:bounds}, we translate these bounds into constraints on the UED parameter space. The black solid lines denote the current IceCube constraint on the UED parameter space for the case of a $B^{(1)}$ (top) or $W^{3(1)}$ (bottom) LKP. These are compared to the current constraints from the CDMS and XENON10 collaborations, as calculated in Ref.~\cite{AKM}. From the top frame of Fig.~\ref{fig:bounds}, we conclude that the current constraints from IceCube are considerably more stringent than those from direct detection experiments for the case of a $B^{(1)}$ LKP. IceCube's constraint in the case of a $W^{3(1)}$ LKP is somewhat less stringent than those obtained from CDMS or XENON10.


To estimate the ultimate reach of IceCube (consisting of a full 80 strings of detectors, and occupying approximately a cubic kilometer of instrumented volume), we have calculated the rate of neutrino-induced muon events from LKP annihilations in the Sun, and compared this to the corresponding rate from atmospheric neutrinos. For this calculation, we have adopted the energy dependent effective area for IceCube as described in Ref.~\cite{GonzalezGarcia:2005xw}, and the spectrum of atmospheric neutrinos given in Ref.~\cite{Honda:2006qj}. Within a $3^o$ window around the direction of the Sun, we estimate that a (completed) IceCube will observe 40 events per year arising from atmospheric neutrinos. This background implies that after three years, IceCube should be able to place an upper limit on the event rate from dark matter annihilations in the Sun of approximately $~6-7$ events per year at the 95\% confidence level. Note that these rates are not per square kilometer of effective area, as are those shown in Fig.~\ref{fig:detrate}, but instead represent the actual rates estimated in IceCube. For a 1 TeV (100 GeV) muon, the effective area of IceCube is approximately 0.75 km$^2$ (0.6 km$^2$). Below $\sim$100 GeV, the effective area of IceCube drops rapidly.


In Fig.~\ref{fig:bounds}, along side the current constraints from IceCube, CDMS and XENON10, we show as dashed lines the projected constraint on the $m_{DM}-r_q$ parameter space plane from the IceCube experiment. For comparison, we also show the projected constraints from the direct detection experiments Super-CDMS and XENON100. 
%
%
 From this figure, we see that for a $B^{(1)}$ LKP, IceCube will likely provide the most stringent constraint, where as in the case of a $W^{3(1)}$ LKP, IceCube's constraint will be comparable to or slightly weaker than those from Super-CDMS and XENON100.





\section{Summary and Conclusions} \label{discussion}

In this article, we have studied the constraints from the high energy neutrino telescope IceCube on Kaluza-Klein dark matter within the context of models with one universal extra dimension. We have considered two Kaluza-Klein dark matter candidates, the first Kaluza-Klein excitations of the hypercharge and isospin gauge bosons, $B^{(1)}$ and $W^{3(1)}$, and calculated the constraints on the UED parameter space from current IceCube data. We have also estimated the future reach of IceCube to such dark matter candidates.  We find that in the case of $B^{(1)}$ dark matter, IceCube currently provides the most stringent constraint on the dark matter mass and its mass splittings with Kaluza-Klein quarks. In the case of $W^{3(1)}$ dark matter, direct detection constraints are slightly stronger than those from neutrino telescopes. IceCube is more sensitive to $B^{(1)}$ annihilations in the Sun than to dark matter most other forms because of its large spin-dependent elastic scattering cross section with nuclei, and its propensity to annihilate to tau leptons or neutrinos.

\emph{Acknowledgements: TF is supported in part by the Federal Ministry of Education and Research (BMBF) under contract number 05H09WWE. TF, AM and KF are supported by the DOE under grant DE-FG02-95ER40899 and by MCTP. DH is supported by the US Department of Energy, including grant DE-FG02-95ER40896, and by NASA grant NAG5-10842.}

\end{document}